\documentclass{aa}
\usepackage{natbib}
\usepackage{txfonts}
\usepackage{graphicx}
\usepackage[figuresright]{rotating}
\bibpunct{(}{)}{;}{a}{}{,}

\begin{document}


\title{Nearby AGN and their hosts in the near infrared\thanks{Based on 
  		observations with the Very Large Telescope of the European 
		Southern Observatory on Paranal, Chile; proposal number 71.B-0062} }

\author{S. Fischer \and C. Iserlohe \and J. Zuther \and T. Bertram \and 
C. Straubmeier \and R. Sch\"odel \and A. Eckart}
\institute{1. Physikalisches Institut, Universit\"at zu K\"oln, Z\"ulpicher Str.
		77, 50937 K\"oln, Germany}
\offprints{S. Fischer, \email{fischer@ph1.uni-koeln.de}}

\abstract{We present near infrared ISAAC VLT observations of nine nearby 
		(0.01$\le$z$\le$0.06) Active Galactic Nuclei
		selected from the Hamburg/ESO Survey and the V\'eron-Cetty \& V\'eron catalog.
		Hydrogen recombination lines Pa$\alpha$ and Br$\gamma$ are observed in seven of the nine sources of which five show a broad component.
		In three sources, extended 1-0S(1) rotational-vibrational molecular hydrogen emission is detected.
		Stellar CO absorption is seen in four sources.
		In one of these objects, an upper limit of the central mass can be determined from the stellar velocity field.
		$H$- and $Ks$-band imaging allow us to determine the morphology class of the host galaxies.
		Colors (with supplementary $J$-band 2MASS images) show that the four galaxies with detected CO absorption are characterized by an overall strong stellar contribution. 
		One galaxy shows an increased extinction towards the nucleus.
		After removal of the nuclear point source, the host galaxies show colors typical for non-active spiral galaxies.
		
		}

\maketitle


\section{Introduction}
  Measuring physical properties of host galaxies of Active Galactic Nuclei (AGN) is mandatory to understand the extreme activity in AGN. 
  Host galaxy properties are under study for some time now. 
  E.g. \citet{jahnke} analyzed mulitcolor data ($BVRi$, $JHKs$) of 19 low redshift quasar hosts. They found that the disc dominated host galaxies show colors similar to average colours of inactive galaxies of the same type, whereas bulge dominated host galaxies appear significantly bluer in ($V-K$) than their inactive counterparts and similar blue as the disc dominated hosts, suggesting a connection between galaxy interaction, induced star formation and the triggering of nuclear activity.
  \citet{reunanen} present NIR-spectroscopy of two Seyfert 1, three Seyfert 2 and one Seyfert 1.5 and show that in Seyfert 2 galaxies Fe II is generally stronger than Br$\gamma$ or H$_2$ lines, while apparently in Seyfert 1 galaxies Br$\gamma$ is stronger.
  \citet{rodriguez2} also studied H$_2$ and Fe II line emission in a sample of 22 (mostly) Seyfert galaxies and concluded that these lines originate in different regions due to systematically different linewidths. They detected molecular Hydrogen in 90\% of their sources, with the majority being excited by thermal excitation processes. They also find a correlation between H$_2$/Br$\gamma$ and Fe II/Pa$\beta$ which can be useful to distinguish emitting line objects by their level of nuclear activity.
  
  In order to study host galaxy properties of QSOs, we created a sample selected from the Hamburg/ESO Survey 
  \citep[HES: e.g.][]{reimers,wisotzki,wisotzki2} and the V\'eron-Cetty \& V\'eron 
  Catalog \citep{veron} with a redshift of z$\leq$0.060.
  This limit grants an observability of the prominent CO(2-0) stellar absorption
  feature in the $K$-band.
  The sample based on these criteria contains 63 sources and represents some of 
  the closest known QSOs for which detailed astrophysical studies (i.e. 
  spatially resolved imaging and spectroscopy) of the host galaxies' physical 
  properties are still possible\footnote{In the following we call this sample the {\it 
  Cologne Nearby QSO Sample}.}.
  We already have successfully performed observations of most of the sources
  in the Cologne Nearby QSO Sample at mm and radio wavelengths with
  telescopes such as BIMA, SEST and the Plateau de Bure interferometer 
  (Bertram et al. in prep., Krips et al. in prep.).
  
  In this paper, we present the first near infrared (NIR) observations on 9 Seyfert 1 galaxies selected from this sample.
  A major advantage of observations in the NIR is that in this wavelength regime extinction is much smaller than in the visible (A$_K=0.112$A$_V$, \citet{rieke}).
  Low resolution (R=500) spectroscopy in the $Ks$-band (2.2$\mu$m) yields
  analysis of several diagnostic lines such as hydrogen recombination lines
  (Pa$\alpha$ \& Br$\gamma$) and rotational-vibrational molecular hydrogen lines.
  The stellar CO absorption bands ($^{12}$CO(2$-$0)) allow analysis of dominating stellar spectral classes and of stellar dynamics.
  Moreover, $J$- (1.25$\mu$m) $H$- (1.65$\mu$m) and $Ks$-band imaging
  provides information on extinction in the galaxy and on whether the nuclear or 
  the stellar component dominates the galaxy's radiation.

  Section 2 of this paper describes the details of the performed NIR 
  observations as well as the data reduction and calibration procedures for
  imaging and spectroscopy.
  In Sect. 3, general spectroscopic results and in Sect. 4 results of the
  photometry are presented.
  Sect. 5 gives a short discussion on the 
  individual sources and is followed by a general summary and conclusion in 
  Sect. 6.

\section{Observation, reduction and calibration}
  9 AGN drawn from the Cologne Nearby QSO sample were observed in seeing
  limited mode with the 
  Infrared Spectrometer and Array Camera (ISAAC) mounted to ANTU (UT1) at ESO's Very Large Telescope (VLT) in Chile during April the 18th-20th 2003.
  ISAAC's 1024$\times$1024 pixel detector provides a pixel scale of 
  0.1484\arcsec/pixel with a field of view (FOV) of 152 $\times$152 arcsec$^2$.
  Our data consist of $H$- and $Ks$-band imaging and $Ks$-band low 
  resolution long slit spectroscopy with a 1\arcsec~slit, resulting in a
  resolution of R=$\lambda/\Delta \lambda$=500.

  The 9 observed sources are selected from the Cologne Nearby QSO Sample to
  comply to observational constraints; all targets are classified as 
  Seyfert 1/narrow line Seyfert 1 galaxies.
  
  Data reduction was carried out with IRAF \& IDL using standard procedures.
  Throughout this paper we use $H_0$=75 km$\,$s$^{-1}\,$Mpc$^{-1}$ and a flat universe with $\Omega_M$=0.3, $\Omega_{vac}$=0.

  \subsection{Imaging}
  
    \begin{table*}
    \caption{Integration times (${\rm t_{int}}$) and seeing conditions for the photometric and 
    	spectroscopic observations. For spectroscopy, also the position angle PA of the 1\arcsec~ slit is listed.
	In all tables and figures throughout this paper, the source 
	\object{VCV(2001) J204409.7-104324} is labeled as VCV J204409.7...}
    \label{DIT}
    \centering
    \begin{tabular}{c|c c c c|c c c}
    \hline\hline
       Source           &\multicolumn{4}{c|}{Imaging}								&\multicolumn{3}{c}{Spectroscopy}\\
       			&\multicolumn{2}{c}{$H$-band}		&\multicolumn{2}{c|}{$Ks$-band}			&\multicolumn{3}{c}{$Ks$-band}\\
			&${\rm t_{int}}$	&Seeing		&${\rm t_{int}}$	&Seeing			&${\rm t_{int}}$	&Seeing		&PA\\
       \hline
       \object{HE 0853-0126}     &7.2s			&1.0\arcsec		&7.2s			&0.9\arcsec			&3900s			&1.8\arcsec	&34$^\circ$\\
       \object{HE 1013-1947}     &7.2s			&1.0\arcsec		&7.2s			&0.8\arcsec			&2700s			&1.6\arcsec	&2$^\circ$\\
       \object{HE 1017-0305}     &8s			&1.0\arcsec		&8s			&0.8\arcsec			&2700s			&0.7\arcsec	&108$^\circ$\\
       \object{HE 1029-1831}     &8s			&0.9\arcsec		&8s			&0.7\arcsec			&2700s			&1.8\arcsec	&-1$^\circ$\\
       \object{HE 1248-1356}     &12s			&1.7\arcsec		&8s			&0.6\arcsec			&2880s			&1.7\arcsec	&103$^\circ$\\
       \object{HE 1328-2508}     &8s			&0.6\arcsec		&8s			&0.5\arcsec			&2880s			&2.7\arcsec	&40$^\circ$\\
       \object{HE 1338-1423}     &8s			&0.5\arcsec		&8s			&0.4\arcsec			&2880s			&0.8\arcsec	&162$^\circ$ (96$^\circ$)\\
       \object{HE 2211-3903}     &8s			&0.6\arcsec		&8s			&0.6\arcsec			&3840s			&0.8\arcsec	&52$^\circ$\\
       VCV J204409.7... &8s			&0.6\arcsec		&8s			&0.7\arcsec			&1440s			&2.6\arcsec	&165.45$^\circ$\\
    \hline
    \end{tabular}
    \end{table*}
  
\begin{figure}
      \centering
      \includegraphics[width=8.5cm]{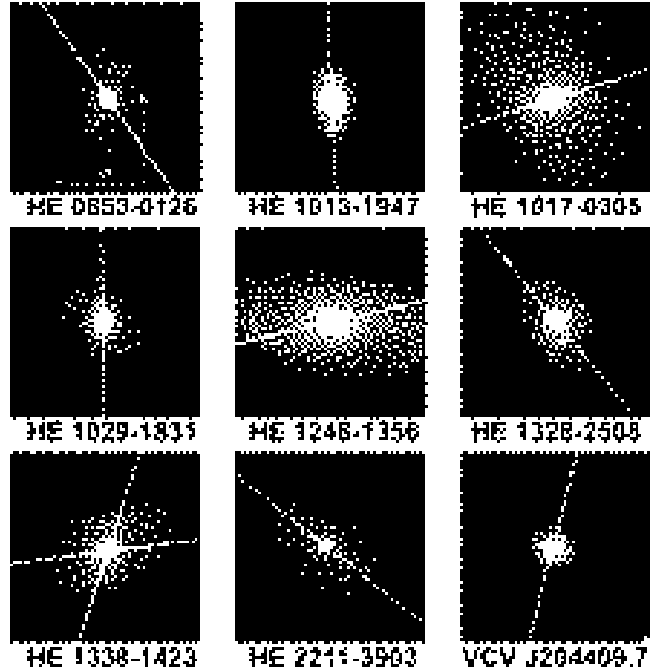}
      \caption{The ISAAC $H$-band images of the 9 observed AGN. Additionally 
      indicated are the slit positions of the spectroscopy run (the slit width
      was 1\arcsec). For HE 1248-1356 the $Ks$-band image is shown.
      }
      \label{images}
    \end{figure}

    The main objective of the observing run was to obtain spectroscopic data,
    therefore only a small fraction of the observation time was assigned to
    imaging.
    The images of each object were taken using position jittering in an ABBA
    source pattern,
    integration times are shown in Tab. \ref{DIT}.
    The background was substracted from the individual frames
    by subtracting consecutive frames from each other.
    After shifting the resulting frames with respect to one reference frame,
    a median image was created.
    Since visual inspection of the flat fields showed that the ISAAC detector
    pixels do not add significant multiplicative signal variations to the images, 
    no flat field was applied.
    The resulting images are shown in Fig. \ref{images}.

    The images were flux calibrated using the data from the 2 Micron All Sky Survey (2MASS).
    With $H$- and $Ks$-band 2MASS-flux values for apertures with a diameter of 14\arcsec (see Tab. \ref{14flux}) zero points (ZP) were calculated for each
    object.
    The H-band 2MASS and ISAAC filter match perfectly.
    Compared to the ISAAC Ks-band filter, the 2MASS filter is slightly broader on the blue side. 
    If the sources show a flat spectrum, the transfer into the 2MASS system does not result in any error. By the means of the slope of our K-band spectra, we estimate that the relative flux error through calibration does not exceed 3\%.
    From the background noise, the fluxcalibration and taking into account that the images were not flatfielded, a conservative error-estimation for H- and K-band fluxes is 10\%.
    In addition to our own ISAAC data we relied on $J$-Band 2MASS images to generate a 2-color diagram.
    Due to high noise in these images, the error is estimated to 15\%.
    Hence the errors for the extracted colors are ${\rm \Delta m}_{H-Ks}{\rm =\pm 0.14^{mag}}$ and ${\rm \Delta m}_{J-H}{\rm =\pm 0.18^{mag}}$.
    
    \begin{table}
    \caption{The 2MASS flux in magnitudes for a 14\arcsec aperture for $J$-, 
    $H$- and $Ks$-band}
    \label{14flux}
    \centering
    \begin{tabular}{c c c c}
    \hline\hline
       Source           &$J$-band  &$H$-band   &$Ks$-band   	\\
    \hline
       HE 0853-0126     &14.02$\pm 0.04$  &13.36$\pm 0.06$  &12.82$\pm 0.07$ 	\\
       HE 1013-1947     &13.53$\pm 0.03$  &12.86$\pm 0.04$  &12.30$\pm 0.04$ 	\\
       HE 1017-0305     &13.29$\pm 0.03$  &12.52$\pm 0.03$  &12.03$\pm 0.04$ 	\\
       HE 1029-1831     &12.30$\pm 0.02$  &12.21$\pm 0.02$  &11.70$\pm 0.03$ 	\\
       HE 1248-1356     &12.24$\pm 0.01$  &11.49$\pm 0.01$  &11.15$\pm 0.02$ 	\\
       HE 1328-2508     &12.17$\pm 0.01$  &11.41$\pm 0.01$  &11.02$\pm 0.01$ 	\\
       HE 1338-1423     &12.65$\pm 0.01$  &11.80$\pm 0.01$  &11.06$\pm 0.02$ 	\\
       HE 2211-3903     &12.69$\pm 0.01$  &11.80$\pm 0.02$  &10.96$\pm 0.01$ 	\\
       VCV J204409.7... &11.74$\pm 0.01$  &10.95$\pm 0.01$  &10.11$\pm 0.01$ 	\\
    \hline
    \end{tabular}
    \end{table}

  \subsection{Spectroscopy}
  \label{specreduc}
    If the host galaxies showed prominent structures already in the acquisition images,
    an alignment of the slit along this structure was preferred.
    The slit angles are indicated in Fig. \ref{images} and presented in Tab. \ref{DIT},
    together with the integration times and the seeing conditions
    for the spectroscopy.
    
    Similar to the imaging, the sky-subtraction was achieved with a
    nodding technique and no flatfielding was applied.
    The individual spectra were corrected for slit-curvature,
    shifting and creating a median then resulted in the final spectrum.
    Telluric correction was performed with B-type stars of which observations
    directly followed or preceded the observations of the targets.
    Hydrogen absorption lines in the telluric star's spectrum were removed 
    by fitting a Lorentz-profile to the line prior to the telluric correction.
    After division by the spectrum of the telluric standard, the galaxy's spectrum
    was multiplied by a blackbody of temperature equal to the effective temperature
    of the star.

    The spectra were wavelength calibrated using OH-lines from the sky.
    Flux-calibration was applied with the zero points calculated from the 
    imaging.
    
    In most of the observations, the seeing was much larger than the used slit width of 1\arcsec (see table \ref{DIT}).
    Hence, the contribution of the AGN to the spectra is diluted and overall the observed spectra will be more influenced by their host-galaxy properties.
    Any flux measurements along the slit depend on the seeing.

\section{Spectroscopy}
\label{Spec2}
  With the integration times given in Tab. \ref{DIT}, we achieved signal-to-noise
  ratios (S/N) on the continuum of 15-40 (the values represent the central 
  spectrum with the highest signal).  

   \begin{figure*}
   \centering
   \includegraphics[width=15cm]{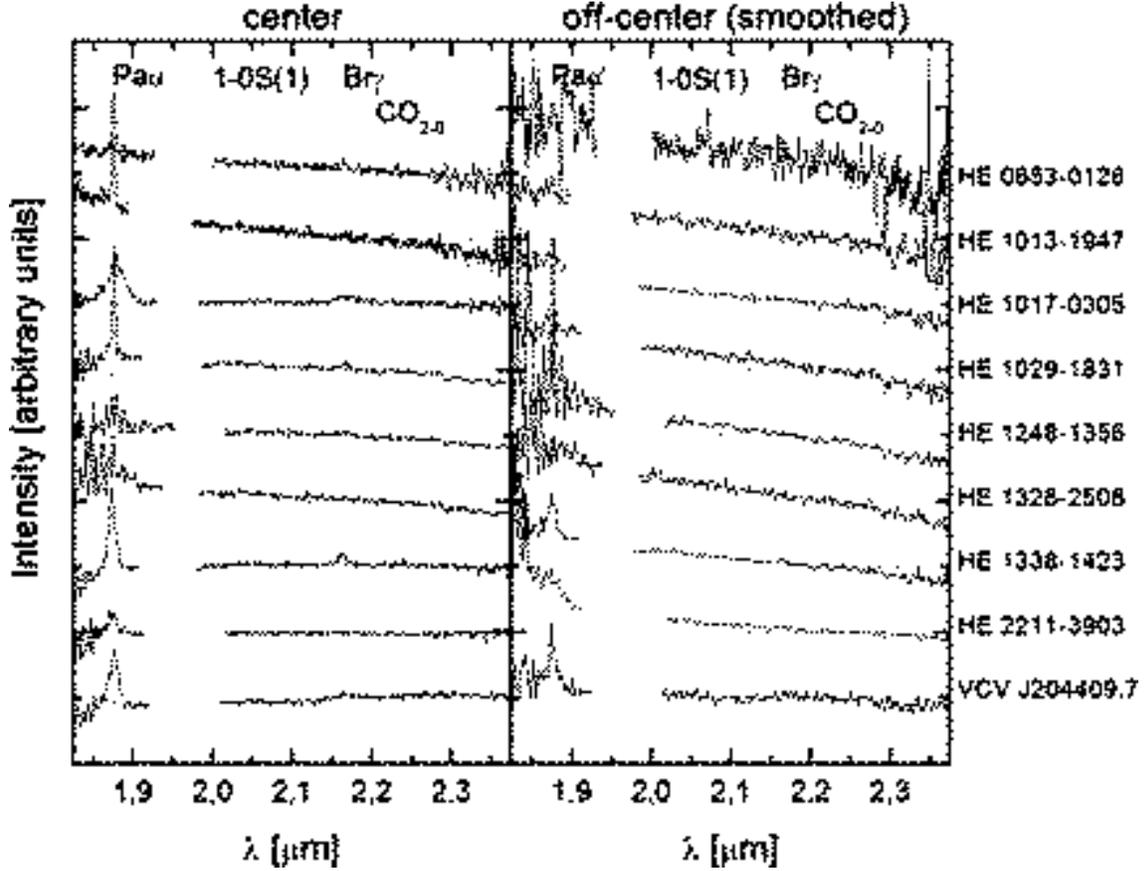}
      \caption{$Ks$-band spectra of the 9 AGN, extracted at the 
      central region (left) and at a distance of 1.5 seeing-FWHM (right, 3-pixel-boxcar-smoothed).
      Indicated are the detected lines in this sample, the Pa$\alpha$ and 
      Br$\gamma$ lines as well as the CO absorption and the H$_2$ 
      (1-0)S(1)-emission line.
      The spectra are presented in restframe and the region at $\sim$2
      $\mu$m is blanked out due to imperfect atmospheric correction.
      The normalization was carried out at 2.2$\mu$m in observer frame.
      Note that the features at  $>2.25\mu$m in HE 0853-0126 probably are not CO absorption features but noise caused by the atmosphere. See section \ref{NoIO} for spectra integrated over the 
      whole galaxy.
      Remarkable are the red continua in most central regions.}
      \label{spektren}
   \end{figure*}
  
  For a comparison between central and off-nuclear spectra, see Fig. \ref{spektren}.
  The spectra are integrated over apertures matching the FWHM of the seeing (see Tab. \ref{DIT}), centered on the nucleus (left) or at a distance of 1.5 times the seeing-FWHM (right).
  Because of low signal to noise in the off-nuclear regions, these spectra are smoothed with a 3-pixel boxcar.
  With respect to the continuum slope, most of the central spectra in Fig. \ref{spektren} show a reddened continuum.
  The CO-detected galaxies appear bluer than the other galaxies (see
  section \ref{COabsorption}).
  However, this effect cannot clearly be seperated from the influences of the seeing (see section \ref{specreduc}).
  
  For details on the analyses of the individual galaxies, as well as spectra integrated over the whole galaxy (i.e. 3 times seeing-FWHM), see section \ref{NoIO}.
  
  \subsection{Hydrogen recombination lines}
  \label{hydrogenrecombination}
  
  \begin{table*}
  	\caption{Hydrogen recombination lines in the 9 AGN. 
	Listed Pa$\alpha$ and Br$\gamma$ fluxes are for spectra integrated over 
	the entire galaxy.
	Calculated extinctions are based on the screen model and following
	an extinction law $\propto \lambda^{-1.75}$.
	The widths of the broad and the narrow component of the Pa$\alpha$ line 
	are for the central spectra.          
	}
	\label{tablehydrogen}      
	\centering          
	\begin{tabular}{c c c c c c}
		\hline\hline
		Source 		&Pa$\alpha$     	&FWHM${\rm _{broad}}$  			& FWHM${\rm_{narrow}}$ 			&Br$\gamma$				&A$_V$		\\ 
    				&$[10^{-23}$\,W\,m$^{-2}]$	&$[$km\,s$^{-1}]$		&$[$km\,s$^{-1}]$			&$[10^{-24}$\,W\,m$^{-2}]$		&$[$mag$]$	\\
		\hline                                                                                                         		                                                        
		HE 0853-0126   &3.3($\pm$3.6\%)  		&				&523($\pm$20\%)				&(4.0($\pm$16.3\%))			&($\geq$0)	\\
    		HE 1013-1947   &13.3($\pm$2.5\%)  	&3190($\pm$20\%)			&437($\pm$20\%)				&7.6($\pm$14.6\%)			&		\\
    		HE 1017-0305   &28.9($\pm$3.3\%) 	&3736($\pm$20\%)			&					&42.7($\pm$5.1\%)			&		\\
    		HE 1029-1831   &56.9($\pm$7.1\%)  	&2081($\pm$20\%)			&92($\pm$40\%)				&49.1($\pm$5.3\%)			&$<$3		\\
    		HE 1248-1356   &  			&					&					&					&		\\
    		HE 1328-2508   &  			&					&					&					&		\\
    		HE 1338-1423   &63.6($\pm$3.5\%)  	&2249($\pm$20\%)			&288($\pm$30\%)				&82.7($\pm$5.0\%)			&		\\
    		HE 2211-3903   &54.0($\pm$8.0\%)  	&	                                &	                                &(21.3($\pm$18.3\%))			&		\\
    		VCV J204409.7... &101.6($\pm$3.1\%)	&2353($\pm$20\%)			&					&101.3($\pm$5.0\%)			&		\\		
		\hline                  
	\end{tabular}
   \end{table*}

    Prominent gaseous diagnostic lines are the hydrogen recombination lines  
    Br$\gamma$ ($\lambda2.1661\mu$m) and Pa$\alpha$ ($\lambda1.8756\mu$m).
    Assuming case B recombination, typical electron densities of 10$^{4}{\rm cm}^{-3}$
    and temperatures of 10000K \citep{osterbrock1989}, line ratios of 
    Pa$\alpha$/Br$\gamma$ are calculated to 12/1.
    We calculated the extinction with an extinction law that follows 
    $\lambda^{-1.75}$ \citep{draine} and using the screen model 
    \citep[e.g.][]{thronson}, values are shown in Tab. \ref{tablehydrogen}.
    
    Pa$\alpha$ and Br$\gamma$ both are detected in seven of the observed 
    galaxies (see Tab. \ref{tablehydrogen}).
    The shape of these lines varies strongly, though.
    The strong Pa$\alpha$ emission allows (with the exception of
    HE 0853-0126 and HE 2211-3903) a separation 
    into a broad and a narrow component.
    However, the broad components of HE 1013-1947 and HE 1029-1831 could
    be influenced by an unresolved He line ($\lambda$1.868$\mu$m), since the
    broad component is rather asymmetric towards the blue.
    
    The broad components of HE 1013-1947 with a FWHM of 3190 km\,s$^{-1} \pm$
    20\% and HE 1017-0305 with a FWHM of 3740 km\,s$^{-1} \pm$20\% 
    show widths characteristic of ordinary Seyfert 1 galaxies.
    In the cases of HE 1029-1831, HE 1338-1423 and VCV(2001) J204409.7-104324, 
    the broad components show widths around 2000 km\,s$^{-1}$, typical for a 
    narrow line Seyfert 1 galaxy \citep[e.g.][]{rodriguez}.

    Only the integrated spectra supplied sufficient S/N to calculate extinction
    in the galaxies within reasonable errors. 
    Case B approximation is only valid for the NLR, but in our Br$\gamma$ lines, no separation between NLR and BLR is possible.
    Although Pa$\alpha$ and Br$\gamma$ are detected in most of the sources, only in HE 0853-0126 and HE 1029-1831, where the line emission is clearly dominated by the NLR, case B is a valid approximation and conclusions on the extinction can be drawn.
    
    In the latter one, visual extinction is smaller than 3$^{mag}$.

    Because of their redshift, the Pa$\alpha$ line in HE 1248-1356 and HE 1328-2508 is shifted into a region of low atmospheric transmission and high variability.
    This could explain why we did not detect this line in these sources.

  \subsection{Molecular hydrogen}
    In three sources, HE 1013-1947, HE 1029-1831 and HE 1328-2508, extended molecular hydrogen is detected in the rotational-vibrational H$_2$ emission line 1-0S(1) at 2.1218$\mu$m (see Figs. \ref{HE1013flux},\ref{HE1029flux},\ref{HE1328flux}).
    It is thought that molecular Hydrogen can be found in 70-80\% of all Seyfert 1 galaxies, \citet{rodriguez2} detected H$_2$ in 90\% of a sample of 22 AGN (19 Seyfert 1).
    Comparingly, with expected H$_2$ (1-0)S(1)/Br$\gamma$ line ratios between 0.6 and 2, our molecular Hydrogen detection rate appears lower.
    But due to the fact that only two of the five galaxies with Br$\gamma$ show also a H$_2$-line, the small sample-size has to be taken into account.
    With the mere detection of the 1-0S(1) emission line, we cannot distinguish between any excitation-mechanisms.

  \subsection{Stellar CO absorption}
  \label{COabsorption}
    The $^{12}$${\rm CO(2-0)}$ ($\lambda 2.295\mu$m) absorption feature of red, evolved stars is by far the strongest absorption feature in the range of 1-3$\mu$m \citep{gaffney1}.
    The absorption is depending on the effective temperature of the star, but also on its luminosity class \citep{kleinmann}.
    Absorption rises towards lower temperatures and from dwarfs to supergiants.
    
    After \citet{doyon} and \citet{goldader}, the spectroscopic CO index,    corresponding to the equivalent width \citep{origlia}, can be used to determine the continuum-dominating stellar luminosity class.
    However, due to the possible contamination of the stellar light by non-thermal radiation, the calculated CO equivalent widths for the AGN in this paper are lower limit estimates for the intrinsic CO band strength.
    In our sample, CO absorption is detected in five AGN at a 2$\sigma$ level with respect to the continuum\footnote{The RMS is determined in the region from (2.00-2.29) $\mu$m (restframe).} (HE 1013-1947, HE 1017-0305, HE 1029-1831, HE 1248-1356 and HE 1328-2508).
    The EWs of the CO absorption are presented in table \ref{COindices}, together with additional upper limits for the remaining sources.
    With the exception of HE 1328-2508, all sources show the described effect that the equivalent width decreases towards the center (but to different extents).
    Therefore, it is difficult to draw conclusions on the underlying stellar composition \citep{origlia2} since it is not clear whether the change in the CO absorption is caused by a rising non-stellar continuum and/or a change in spectral class.
    Assuming that the values deduced in the outer regions of the galaxy are rather free from non-stellar continuum, a depth of (6 $\pm$ 3)~\AA~ as it can be found in HE 1029-1831 and HE 1328-2508 corresponds to the value expected for that of K0-3 giants \citep{kleinmann}. The comparatively deep CO-absorption in HE 1248-1356 with a width of (11 $\pm$ 2)~\AA~  points to late K or early M giants. 
    In starburst galaxies, deep CO absorption is linked to young red supergiants.
    A typical starburst galaxy such as M82 shows an equivalent width of (15 $\pm$ 1)~\AA~ \citep{tamura} which is significantly deeper than found in HE 1013-1947, HE 1029-1831 and 1328-2508. 
    Their values are consistent with those found in ordinary elliptical and spiral galaxies \citep{frogel}.
    Therefore, the absorption in HE 1248-1356 can most likely be attributed to ongoing starformation in its host galaxy.
    A narrow EW of $\sim$3~\AA~ as it can be found in the host of HE 1017-0305 is typical for G/K dwarfs \citep{kleinmann}.
    In HE 0853-0126, the absorption at wavelengths $>$2.25 $\mu$m is probably caused by atmospheric noise. 
    The line-shape does not resemble that of stellar CO-absorption.
    
    \begin{table}
    \caption{The CO equivalent width in~\AA~ for the four sources with detected stellar CO absorption. The values in italics are upper limits.}
    \label{COindices}
    \centering
    \begin{tabular}{c c c}
    \hline\hline
       Source          &Central EW$_{CO}$	&Off-nuclear CO EW$_{CO}$\\
    \hline
       HE 0853-0126    &(\textit{7})			&(3 $\pm$ 1)\\
       HE 1013-1947    &4 $\pm$ 2			&7 $\pm$ 4	\\
       HE 1017-0305    &1.2 $\pm$ 0.2			&3 $\pm$ 1 \\
       HE 1029-1831    &5 $\pm$ 2			&6 $\pm$ 3	\\
       HE 1248-1356    &4 $\pm$ 1			&11 $\pm$ 2	\\
       HE 1328-2508    &6 $\pm$ 2			&6 $\pm$ 3	\\
       HE 1338-1423    &\textit{1.3}			&\textit{4.9}\\
       HE 2211-3903    &\textit{1.2}			&\textit{1.5}\\
       VCV J204409.7...&\textit{2.8}			&\textit{1.8}\\
    \hline
    \end{tabular}
    \end{table}

      \begin{figure}
        \centering
        \includegraphics[width=8.5cm]{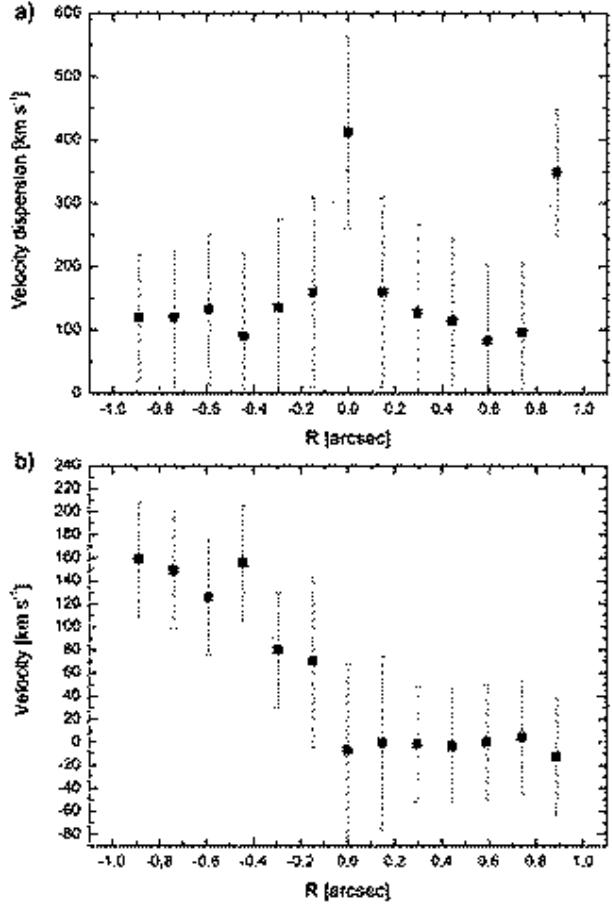}
        \caption{HE 1248-1356: In a) the stellar velocity dispersion 
	and in b) the stellar rotation curve is shown. 
	In the central spectrum, the velocity dispersion
	is calculated to 411$\pm$150 km\,s$^{-1}$, see text for details.}
        \label{rot+disp}
      \end{figure}

    The CO(2-0) bandhead is also a useful feature for investigations of the underlying stellar dynamics.
    The observed CO absorption in the galaxy is (in a rough approximation) the convolution of the star that dominates the CO absorption with the line-of-sight velocity profile (LOSVP), so the LOSVP can either be determined by basic deconvolution processes, or via comparison to a convolved spectrum of a star or a stellar population.
    For best results with our galaxy spectra of low resolution and S/N ratios, we fitted a stellar template spectrum, convolved with a Gaussian distribution (representing the LOSVP), to the galaxy spectrum \citep{gaffney1} with the Nelder-Mead downhill simplex algorithm.
    Note that in this procedure, the selection of the correct stellar template is essential since template-mismatch produces significantly too high velocity dispersions.
    The characteristic for the selection of the stellar spectrum is an according equivalent width of the CO(2-0) absorption bandhead to the galaxy's spectrum \citep{gaffney2}.
    Hence our deduced velocity dispersion is an upper limit to the intrinsic velocity dispersion of the galaxy.
    For the analysis in this sample, we relied on spectra of K- and M- giants and supergiants in the $K$-band from \citet{hinkle} (native resolution of R=3000)\footnote{These sources can be found online in the SIMBAD database, which is operated at CDS, Strasbourg, France; url: http://simbad.u-strasbg.fr/Simbad.}.
    Convolution with a Gaussian distribution result in K-band stellar template spectra at a resolution of R=500.
    
    Among the sources with detected CO absorption, HE 1248-1356 is the only object in our sample where the stellar velocity  field is resolved.
    Fig. \ref{rot+disp}a) and \ref{rot+disp}b) show the resulting velocity dispersion and stellar rotational curve of HE 1248-1356.
    The seeing during the observations of HE 1248-1356 was 1.7\arcsec~ (see Tab. \ref{DIT}). 
    The deduced velocity dispersion of $\sigma$=400 $km\,s^{-1}$ yields an enclosed virialmass of 2$^.10^{10}$ M$_\odot$ in the seeing disc (= 0.5 kpc), a typical value for Seyfert galaxies.
    However, this mass estimate has to be taken with caution because the seeing during observations was large and the shown data-points are strongly correlated. Depending on the exact shape of the rotation curve and the distribution of the velocity dispersion, $\sigma$ is likely to increase under better seeing conditions (i.e. smaller R). Here clearly investigations under higher angular (and spectral) resolution are required.
    
\section{Photometry}
\label{photom2}

    \begin{figure*}
    \sidecaption
      \centering
	   \includegraphics[width=14cm]{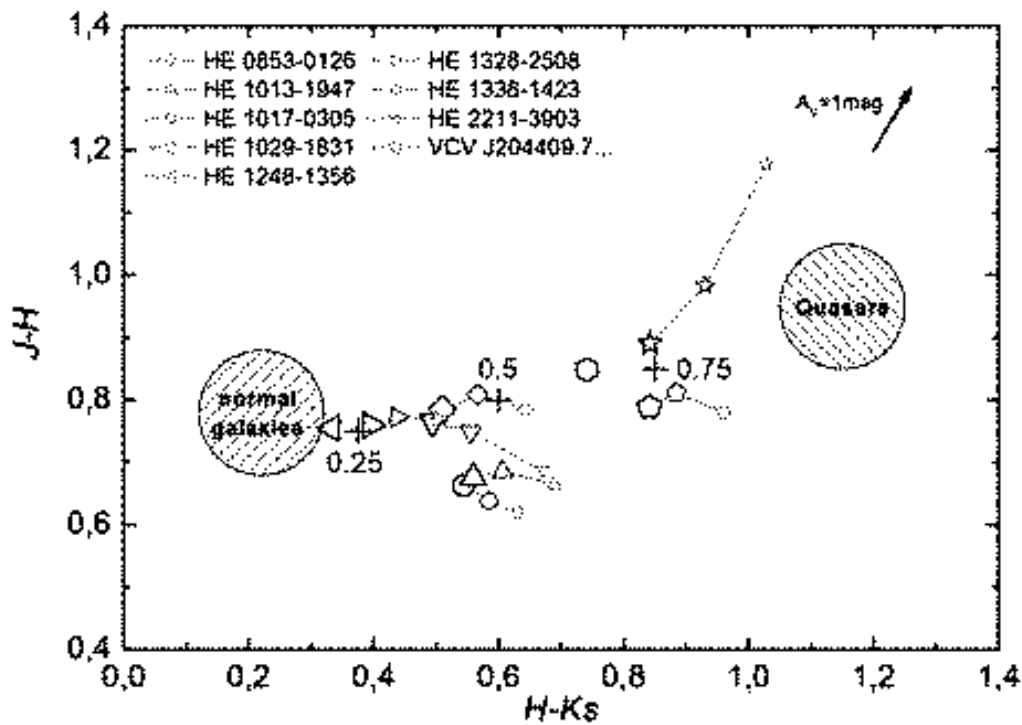}
	   \caption{The NIR 2-color diagram for the 9 Seyfert 1 galaxies.
               The three different sized data markers for each galaxy represent
	       the flux measurements in apertures of 14\arcsec, 8\arcsec and
	       3\arcsec, centered on the nucleus.
	       Additionally, the regions of normal 
	       galaxies and Quasars and 
	       the arrow specifying a visual extinction of 1${\rm ^{mag}}$ 
	       are indicated in the graph.
	       The locus of the three crosses (in grey) is based upon 
	       calculations by \citet{hyland} and refers to the radiation 
	       output from normal 
	       galaxies with gradually increased Quasar contribution (+$_{0.25}$: 
	       25\% nuclear radiation...).}
      \label{2color}
    \end{figure*}

    \begin{table}
  	\caption{Morphology and redshifts. 
	The morphological classifications are deduced in this paper.
	The references for the redshifts are as follows: $[1]$: LEDA/SIMBAD 
	2002 comparison, $[2]$: QDOT all-sky IRAS galaxy redshift survey,
	$[3]$: \citet{wisotzki2}, $[4]$: NED team.
	}
	\label{literature}      
	\centering          
	\begin{tabular}{c c c c}
		\hline\hline
		Source 		&morphological     	&redshift		&references\\   
    				&classification		&			&\\             
		\hline                                                                          %
		HE 0853-0126   &SBb/c			&0.059903		&$[1]$\\        
    		HE 1013-1947   &SB			&0.055			&$[3]$\\        
    		HE 1017-0305   &interacting (S)		&0.0468			&$[4]$\\        
    		HE 1029-1831   &SBc			&0.040134		&$[1]$\\        
    		HE 1248-1356   &Sa/SBa			&0.014557		&$[1]$\\        
    		HE 1328-2508   &interacting (S)		&0.026485		&$[1]$\\        
    		HE 1338-1423   &SB			&0.041752		&$[1]$\\        
    		HE 2211-3903   &SB			&0.038473	        &$[2]$\\        
    		VCV J204409.7... &E3			&0.034373		&$[2]$\\	
		\hline                                                                                                  
	\end{tabular}
   \end{table}

  ISAAC $H$-band images of each source are presented in Fig. 
  \ref{images}.
  The images clearly resolve structures of the underlying host galaxy like 
  bars and spiral arms, even without a subtraction of the nucleus; for a 
  morphologic classification of the individual host galaxies, see Tab. \ref{literature}.
  The sample is mostly dominated by spiral galaxies.
  Since all our galaxies are lower luminosity AGN, this supports the theory that
  the propability to find disc dominated host galaxies increases towards lower 
  luminosity nuclei \citep[e.g.][~and references therein]{jahnke}.
  For more details on the morphology, see Sect. \ref{NoIO}.

  In combination with the J-band data from 2MASS we derived colors for each
  of the 9 AGN in apertures centered on the nucleus.
  In the NIR, colors of an active galaxy consist of several components:
  A stellar component, a Quasar-like component, an extinction-component and a
  component associated to hot dust.
  \citet{hyland} calculated mean colors of Quasars to $(J-H)=0.95$ and
  $(H-Ks)=1.15$ (corrected for redshift), while 
  ordinary galaxies have colors of $(J-H)=0.78$, $(H-Ks)=0.22$
  \citep{glass1}.
  Wavelength dependent extinction is expressed by a shift along the direction 
  of a vector representing a visual extinction of ${\rm 1^{mag}}$ (following
  $A_V:A_J:A_H:A_K=1:0.282:0.175:0.112$ \citep{rieke,binney}).
  Re-radiation of hot dust at temperatures of $\sim$300 K in the surroundings
  of the nucleus results in an increased flux especially in the K-band, so
  in the 2-color diagram this effect manifests itself in a horizontal shift
  in $H$-$Ks$ \citep{glass1}.
  
  We measured the flux of each galaxy in apertures of three different sizes,
  always centered on the nucleus:
  One aperture size basically encloses the whole object (14\arcsec ~diameter),
  one encloses only the nucleus (3\arcsec) and one intermediate size
  (8\arcsec) was chosen.
  A K-correction was not applied because at the reshifts of the sample its 
  effect is contained in the photometric errorbars.
  To compare the ISAAC seeing limited images (resolution: (0.6-1)\arcsec) 
  to the 2MASS $J$-band images (resolution: (2.3-2.8)\arcsec) the ISAAC 
  images were smoothed with a Gaussian
  distribution such that the $H$- and $Ks$-band images show resolutions
  corresponding to their respective $J$-band images.
  The smallest used aperture measures 3\arcsec, according to the resolution 
  limit of the 2MASS data.
  The resulting 2-color diagram is shown in Fig. \ref{2color}.
  For HE 1338-1423 and HE 1248-1356\footnote{Due to unfortunate
  shifting in the observations of HE 1248-1356, no H-band image could be
  generated. Here we additionally relied on a 2MASS H-band image.} only the 
  color value extracted from the largest aperture is shown because the $J$-band 
  colors in smaller apertures do not provide physical results.
  
  \begin{table}
  	\caption{Measured $JHKs$ magnitudes of the aperture photometry in the 8\arcsec 
	(first line) and 3\arcsec (second line)
	aperture (since the photometric calibration was performed in reference to the
	2MASS 14\arcsec~ aperture data, according magnitudes can be found in 
	Tab. \ref{14flux}). In addition, the $H-Ks$-colors of the host galaxy 
	are presented (see text for details on the process of subtraction of the 
	nucleus).}
	\label{apertphot}
	\centering
	\begin{tabular}{c c c c c}
		\hline\hline
		Source 		&$J$-band   	&$H$-band	&$Ks$-band	&$H-Ks$\\
		\hline
		HE 0853-0126   	&14.451		&13.813		&13.228		&0.31\\
		 		&15.721		&15.102		&14.470\\
		HE 1013-1947	&13.838		&13.153		&12.547		&0.51\\
		 		&15.005		&14.342		&13.653\\
		HE 1017-0305    &13.690		&12.941		&12.386		&0.55\\
		 		&15.002		&14.315		&13.646\\
		HE 1029-18311   &13.232		&12.422		&11.856		&0.31\\
		 		&14.337		&13.551		&12.909\\
		HE 1248-1356    &(12.765)	&(12.853)	&(11.601)	&-\\
		 		&(14.132)	&(14.588)	&(12.856)\\
		HE 1328-2508    &12.551		&11.778		&11.341		&0.28\\
		 		&13.865		&13.094		&12.591\\
		HE 1338-1423    &(12.950)	&(12.189)	&(11.364)	&0.48\\
		 		&(14.120)	&(13.588)	&(12.671)\\
		HE 2211-3903    &12.998		&12.016	        &11.086		&0.34\\
		 		&14.220		&13.041		&12.012\\
		VCV J204409.7... &11.908	&11.097		&10.212		&1.20\\
		 		&12.934		&12.154		&11.193\\
		\hline
	\end{tabular}
  \end{table}
  
  As expected for Seyfert 1 galaxies \citep[cf.][]{ward} the global colors of the
  9 sources are located in between the regions of Quasar colors and colors of
  ordinary galaxies.
  They all are distributed along the line connecting the two extreme cases, thus
  indicating a varying importance of the Quasar component in each object.
  The colors of HE 2211-3903 and VCV(2001) J204409.7-104324
  show that these galaxies are dominated by their non-stellar nucleus.
  A dominating stellar component in HE 1248-1356, HE 1328-2508,
  HE 1013-1947 and HE 1029-1831 could be connected with the
  detection of stellar CO absorption in the spectra of exactly these sources,
  because the strong contribution of the non-stellar nuclei in the other
  sources can prevent such detection.
  The continuum slopes of the K-band spectra demonstrate similar tendencies.
  
  Rising influences of the nucleus are expressed in a shift of the color
  towards the region of Quasars with decreasing aperture size.
  This effect is superposed onto the effect of hot dust.
  All targets show increased nuclear radiation (in comparison to the stellar 
  radiation) or reradiation of hot dust but no sigificant amounts of extinction
  towards the nucleus, with the exception of
  HE 2211-3903 with a visual extinction of ${\rm \sim 2.5^{mag}}$.
  Especially, HE 1017-0305 does not stand out with extraordinary
  amounts of extinction as the spectroscopically determined A$_V$ would imply,
  but the photometric colors are averaged over larger apertures.
  
  We also estimated H-K colors of the host galaxies by subtracting the 
  contribution of the nucleus.
  Here we used the fact that the nucleus is unresolved, i.e. star like in extent.
  Hence the subtraction was performed by taking a star in the vicinity of the 
  galaxy, shifting it to the center of the nucleus within sub-pixel accuracy 
  and subtracting it scaled to the flux of the nucleus such that just no "hole"
  is produced in the galaxy.
  Ideally, this results in a smooth brightness distribution representing
  the contribution of the host galaxy.
  The flux of the host was determined in a 14\arcsec aperture, values are shown
  in the last column of Tab. \ref{apertphot}.
  The spiral galaxies show host-colors similar to or slightly redder than 
  the colors of non-active spirals.
  The color of the elliptical galaxy VCV(2001) J204409.7-104324 is 
  probably influenced by difficulties in removing the nucleus in the K-band.
  For HE 1248-1356 no host colors were calculated due to the lack of
  a high-resolution H-band image.

\section{Notes on individual objects}
\label{NoIO}
\begin{figure}
  \centering
  \includegraphics[width=8.5cm]{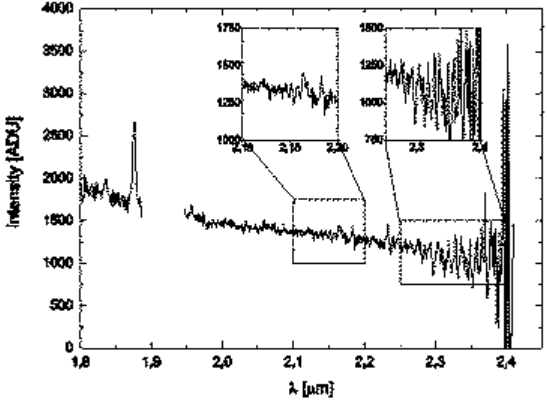}
      \caption{The spectrum of HE 0853-0126, integrated over 5.3\arcsec (= 5 kpc). 
      In addition to the overall K-band spectrum, the region of the Br$\gamma$
      line and the CO absorption are shown.}
      \label{HE0853}
   \end{figure}

\begin{figure}
   \centering
   \includegraphics[width=8.5cm]{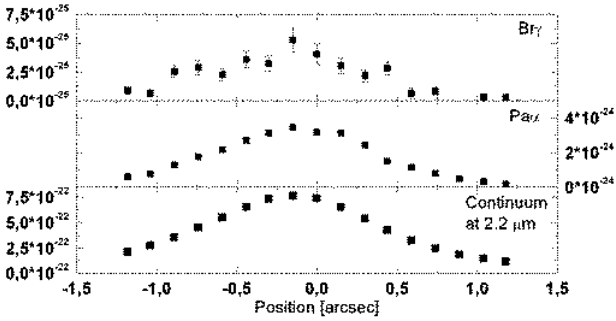}
      \caption{Flux of HE 0853-0126. 
      The flux of the Pa$\alpha$ and Br$\gamma$
      line is given in units of W\,m$^{-2}$, the continuum flux is in units of 
      W\,m$^{-2}$\,$\mu$m$^{-1}$. 
      Where no errorbars can be seen, the data-marker's size exceeds the 
      extension of the errorbars.}
      \label{HE0853flux}
   \end{figure}

\begin{figure}
   \centering
   \includegraphics[width=8.5cm]{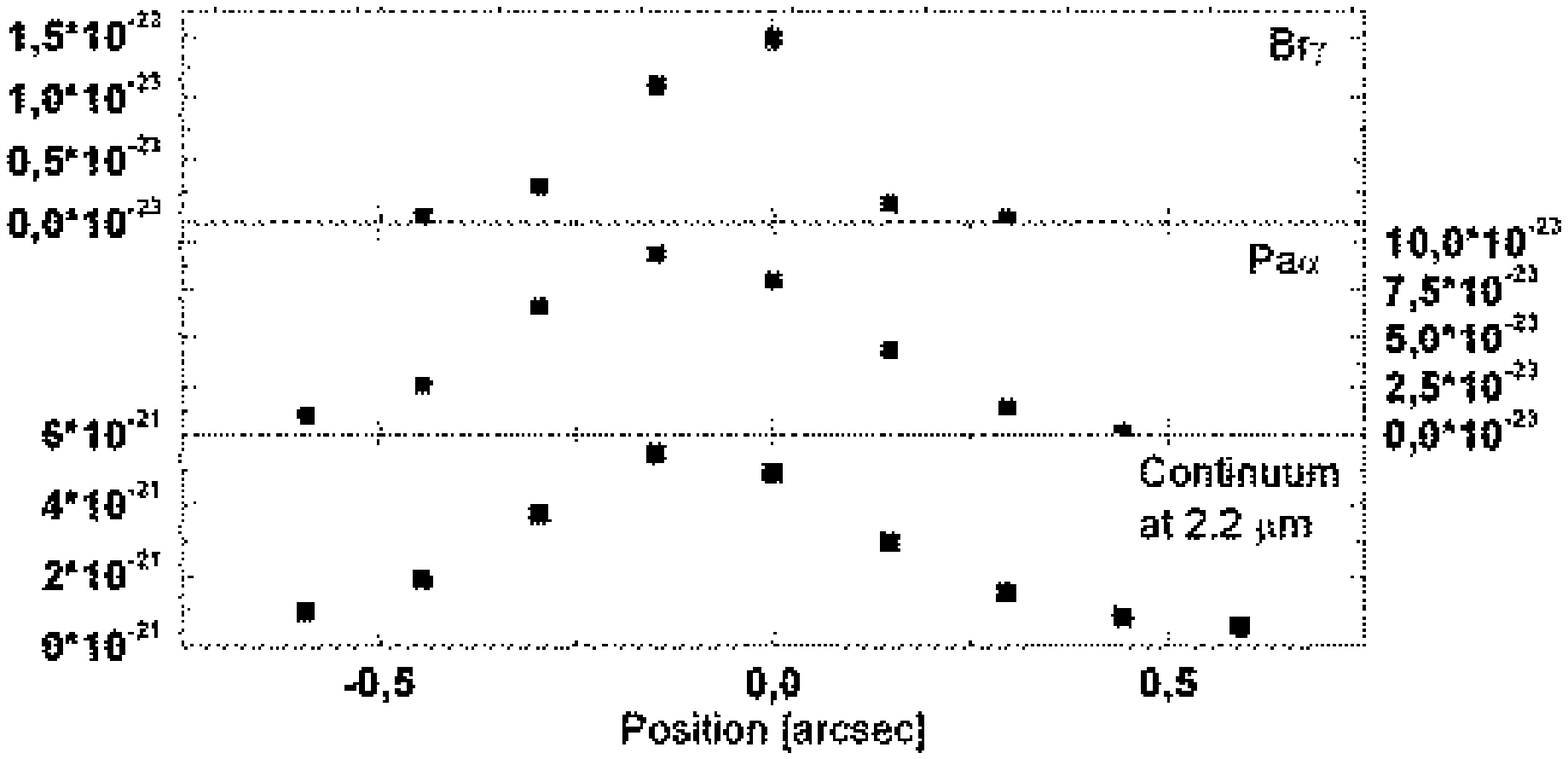}
      \caption{Flux of HE 1017-0305. 
      The flux of the Pa$\alpha$ and Br$\gamma$
      line is given in units of W\,m$^{-2}$, the continuum flux is in units of 
      W\,m$^{-2}$\,$\mu$m$^{-1}$.}
      \label{HE1017flux}
   \end{figure}

\begin{figure}
  \centering
  \includegraphics[width=8.5cm]{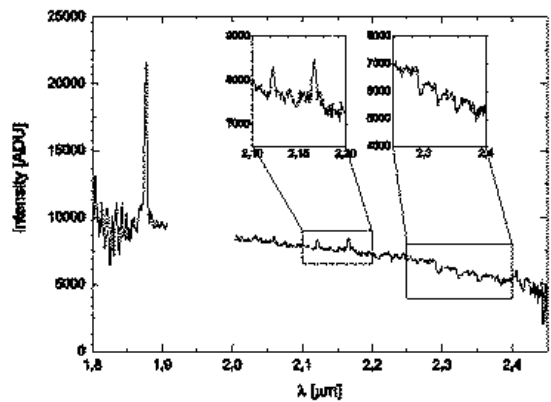}
      \caption{The over 5.3\arcsec (= 7.3 kpc) integrated spectrum of HE 1029-1831. 
      In addition to the overall K-band spectrum, the region of the 1-0S(1)H$_2$ line
      and the Br$\gamma$ line as well as the CO absorption are shown.}
      \label{HE1029}
   \end{figure}

\begin{figure}
  \centering
  \includegraphics[width=8.5cm]{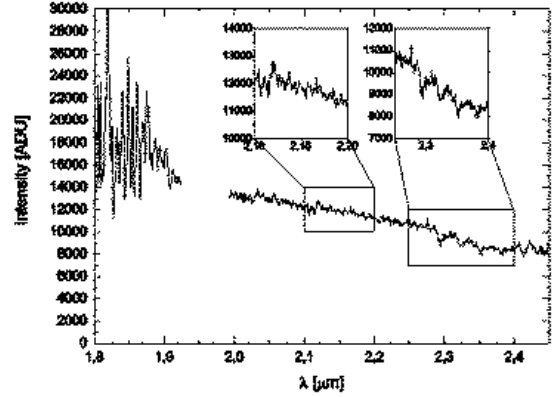}
      \caption{The over 8.0\arcsec (= 16.2 kpc) integrated spectrum of HE 1328-2508. 
      In addition to the overall K-band spectrum, the region of the Br$\gamma$ line as well as the CO absorption are shown.}
      \label{HE1328}
  \end{figure}

\begin{figure}
   \centering
   \includegraphics[width=8.5cm]{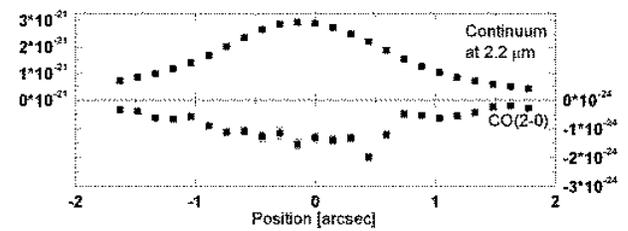}
      \caption{Flux of HE 1328-2508. 
      The flux of the  ${\rm CO(2-0)}$ flux deficit is given in units of W\,m$^{-2}$, 
      the continuum flux is in units of W\,m$^{-2}$\,$\mu$m$^{-1}$.}
      \label{HE1328flux}
   \end{figure}

\begin{figure}
  \centering
  \includegraphics[width=8.5cm]{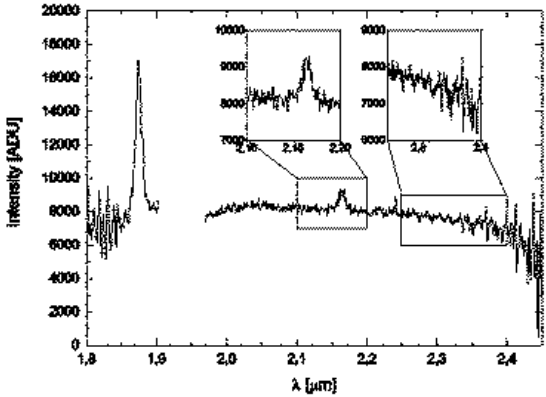}
      \caption{The over 2.5\arcsec (= 3.3 kpc) integrated spectrum of HE 1338-1423. 
      In addition to the overall K-band spectrum, the region of the Br$\gamma$ line as well as the CO absorption are shown.}
      \label{HE1338}
  \end{figure}

\begin{figure}
   \centering
   \includegraphics[width=8.5cm]{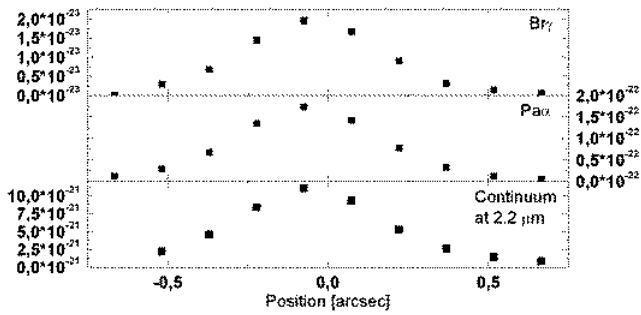}
      \caption{Flux of HE 1338-1423. 
      The flux of the Pa$\alpha$ and the Br$\gamma$
      line is given in units of W\,m$^{-2}$, 
      the continuum flux is in units of W\,m$^{-2}$\,$\mu$m$^{-1}$.}
      \label{HE1338flux}
   \end{figure}

\begin{figure}
  \centering
  \includegraphics[width=8.5cm]{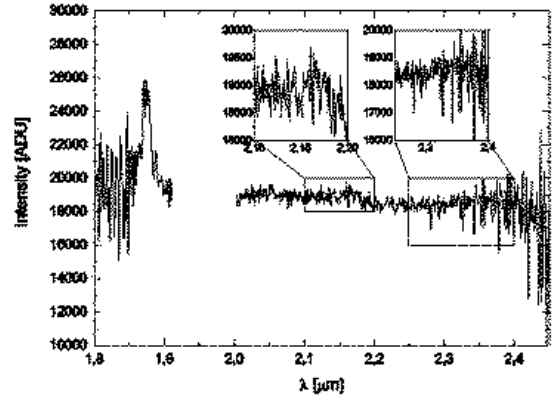}
      \caption{The over 2.5\arcsec (= 3.6 kpc) integrated spectrum of HE 2211-3903. 
      In addition to the overall K-band spectrum, the region of the Br$\gamma$ line as well as the CO absorption are shown.}
      \label{HE2211}
  \end{figure}

\begin{figure}
   \centering
   \includegraphics[width=8.5cm]{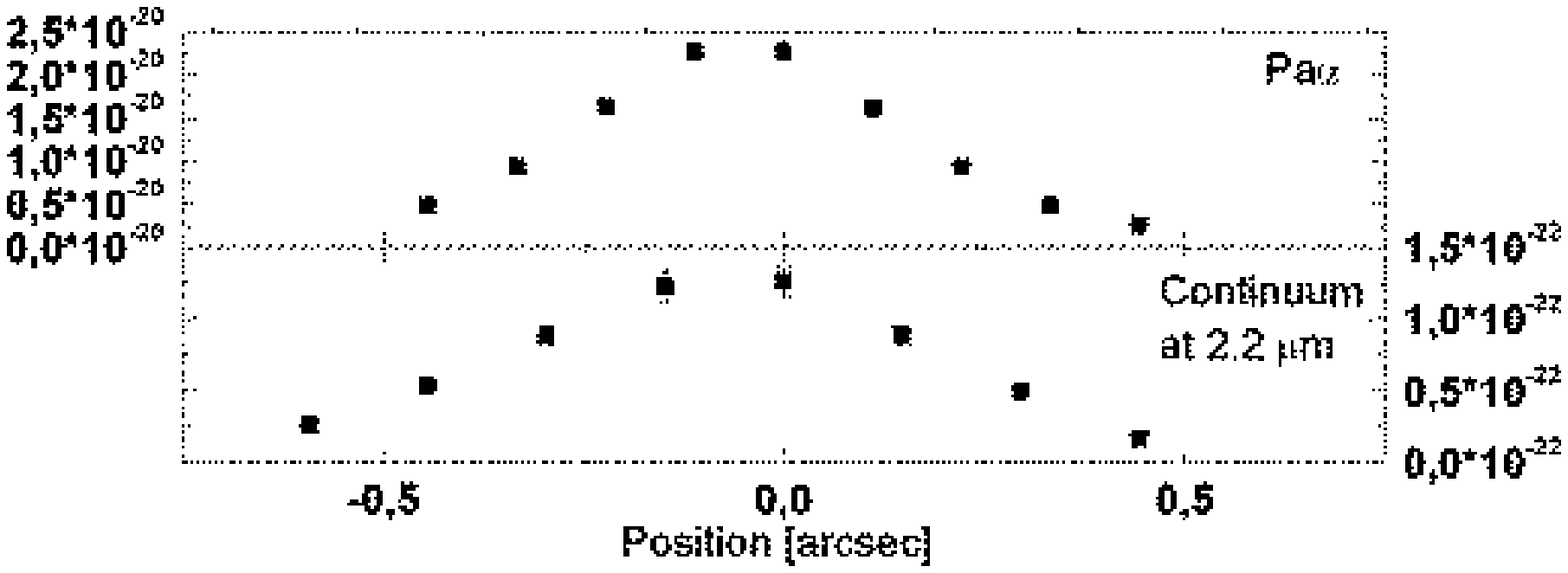}
      \caption{Flux of HE 2211-3903. 
      The flux of the Pa$\alpha$
      line a is given in units of W\,m$^{-2}$, 
      the continuum flux is in units of W\,m$^{-2}$\,$\mu$m$^{-1}$.}
      \label{HE2211flux}
   \end{figure}

\begin{figure}
  \centering
  \includegraphics[width=8.5cm]{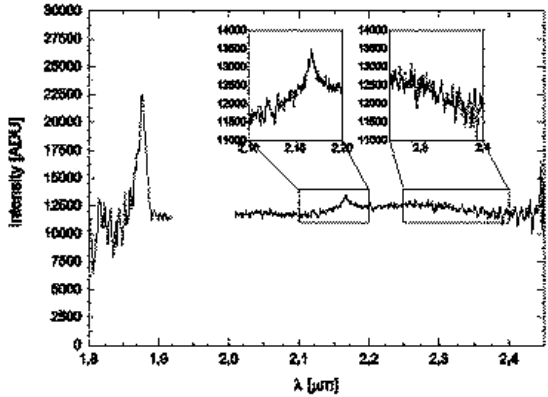}
      \caption{The over 7.7\arcsec (= 12.2 kpc) integrated spectrum of VCV(2001) J204409.7-104324. 
      In addition to the overall K-band spectrum, the region of the Br$\gamma$ line as well as the CO absorption are shown.}
      \label{VCVJ20}
  \end{figure}

\begin{figure}
   \centering
   \includegraphics[width=8.5cm]{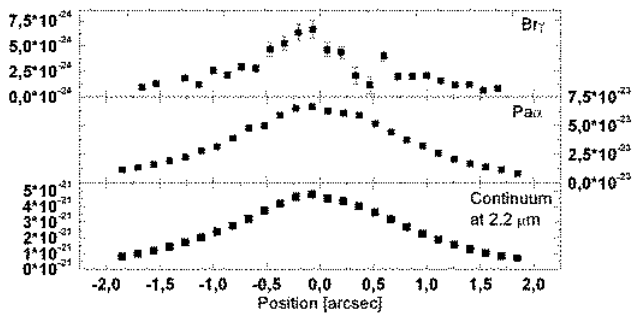}
      \caption{Flux of VCV(2001) J204409.7-104324. 
      The flux of the Br$\gamma$ and the Pa$\alpha$
      line is given in units of W\,m$^{-2}$, 
      the continuum flux is in units of W\,m$^{-2}$\,$\mu$m$^{-1}$.}
      \label{VCVJ20flux}
   \end{figure}

The spectra, integrated over 3 times the seeing-FWHM, the line fluxes and the flux of the continuum of the objects, extracted along the slit, are presented in Fig. \ref{HE0853}-\ref{VCVJ20flux}.
In addition, a short discussion of each galaxy is given.
Note that in most cases, the seeing during the spectroscopic observations was larger than the slitwidth (c.f. section \ref{specreduc}).

  \subsection{HE 0853-0126}
  \label{apHE0853}

    HE 0853-0126 shows a prominent bar structure (the slit was not placed
    along this bar because the structure was too faint to be resolved during acquisition). 
    Two spiral arms can be seen reaching symmetrically out to the east and the 
    west.
    The view on the galaxy is rather face-on.
    The region of the CO absorption is redshifted to the long wavelength cutoff of the K-band. In the resulting high noise, no CO absorption can be detected.
    Hydrogen recombination is seen in Pa$\alpha$ and Br$\gamma$,
    a broad component in the Pa$\alpha$ line is not detected, the
    galaxy is classified as Seyfert 1 in \citet{wisotzki2}.
    A weak broad component could either be hidden in the noise of the spectrum, or the AGN is variable.
    In the region of (2.0-2.2)$\mu$m, the spectrum shows a S/N of 21 on the continuum.
    This causes high errors in the Br$\gamma$ flux measurements and in the
    calculated visual extinction.

  \subsection{HE 1013-1947}
  \label{apHE1013}
  
  \begin{figure}
  \centering
  \includegraphics[width=8.5cm]{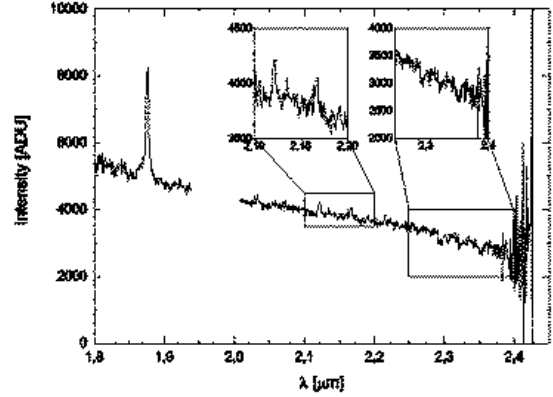}
      \caption{The over 4.9\arcsec (= 5 kpc) integrated spectrum of HE 1013-1947. 
      In addition to the overall K-band spectrum, the region of the 1-0S(1)H$_2$ line
      and the Br$\gamma$ line as well as the CO absorption are shown.}
      \label{HE1013}
   \end{figure}

   \begin{figure}
   \centering
   \includegraphics[width=8.5cm]{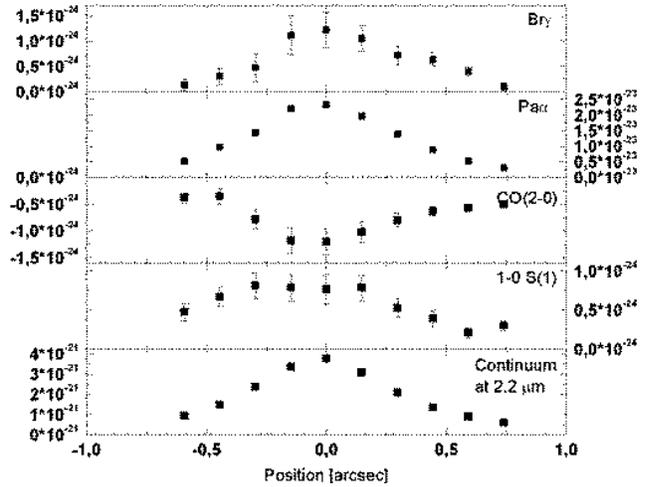}
      \caption{Flux of HE 1013-1947. 
      The flux of the Pa$\alpha$, Br$\gamma$, (1-0)S(1) H$_2$-line
      and the the flux deficit in ${\rm CO(2-0)}$
      is presented in units of W\,m$^{-2}$, the continuum flux is in units of 
      W\,m$^{-2}$\,$\mu$m$^{-1}$.}
      \label{HE1013flux}
   \end{figure}
   
  The galaxy is a barred spiral galaxy and shows a very bright nucleus.
  \citet{nagao1} as well as \citet{rodriguez} find HE 1013-1947 to be a narrow line Seyfert 1 galaxy with a broad component in H$\alpha$ and H$\beta$ of about 1900 km\,s$^{-1}$, which is narrower than the line width we extracted using Pa$\alpha$.s
  The difference between the optical and NIR-linewidths could be explained by extinction effects.
  The slight blueshift of the center of the broad component  versus the center of the narrow component could indicate that an unresolved He line ($\lambda$1.838$\mu$m) influences our measurement.
  
  The integrated spectrum in Fig. \ref{HE1013} also shows CO absorption
  at wavelengths $\ge$2.29 $\mu$m.
  In the central spectrum, the equivalent width of the ${\rm CO(2-0)}$ absorption is
  (3.7 $\pm$ 2.4)~\AA, rising to an equivalent width of (7.5 $\pm$ 4)~\AA~in the 
  outer regions of the galaxy (see Tab. \ref{COindices}).
  The equivalent width at the center of the galaxy is diminished due to a larger contribution of the non-thermal continuum at the nucleus (the overall absorption in the ${\rm CO(2-0)}$ transition rises towards the nucleus, as can be seen in Fig. \ref{HE1013flux}).
  An equivalent width of 7~\AA~ corresponds to the value found for that of K0-3 giants \citep{kleinmann}.
  
  The galaxy also shows extended H$_2$ emission in the (1-0)S(1) transition
  with a total flux of (7.4 $\pm$ 1)$^.10^{-24}$\,W\,m$^{-2}$,
  the line ratio to the Br$\gamma$ line is calculated to 1.0 $\pm$ 0.2.
  According to \citet{rodriguez2}, this ratio is a typical value for a Seyfert galaxy.
  Since this is the only detected molecular hydrogen line, excitation 
  processes cannot be discussed.

  \subsection{HE 1017-0305}
  \label{apHE1017}
   
  \begin{figure}
  \centering
  \includegraphics[width=8.5cm]{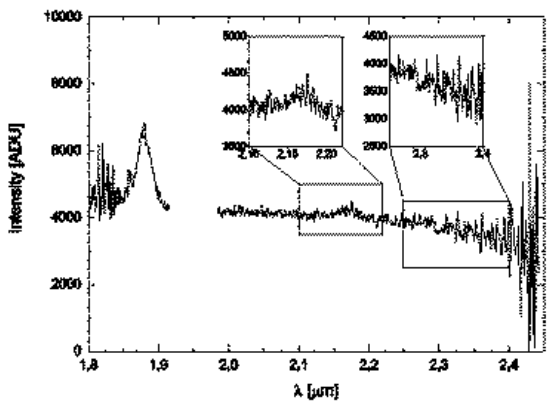}
      \caption{The over 2.2\arcsec (= 2.6 kpc) integrated spectrum of HE 1017-0305. 
      In addition to the overall K-band spectrum, the region of the Br$\gamma$
      line and the CO absorption are shown.}
      \label{HE1017}
   \end{figure}
  
  The radio quiet \citep{wadekar} galaxy shows a very prominent bar extending 
  from east to west.
  Spiral arms are also resolved, but their shape is rather unsymmetric
  which implies a disturbance possibly caused by a recent merger event.
  At a distance of 42\arcsec~ to the NW of the object, the galaxy LEDA 1072782 
  with a redshift of z=0.04925 \citep{updatedzwicky} can be found
  which supports the idea of an ongoing merger event.
  In Fig. \ref{spektren}, the continuum slopes at the center and away
  from the nucleus change significantly, with the spectrum beeing
  much bluer towards larger distances from the nucleus.
  This is also represented by the calculated extinction of (16 $\pm$ 19)$^{mag}$,
  but the high noise on the broad Br$\gamma$ line (see Fig. \ref{HE1017}) produces large errors in the measurement of its flux.
  In the region between 2 $\mu$m and 2.2 $\mu$m, the S/N on the continuum is 40.
  The two broad hydrogen recombination lines make HE 1017-0305 a typical representative
  for a Seyfert 1 galaxy.
  In the 2-color diagram (Fig. \ref{2color}), the galaxy shows no strong amounts of reddening (note that due to the 2MASS-limited resolution, the extinction is derived from a larger area).
  
  The upper limits for the IRAS fluxes (IRAS Vol. II catalog) yield an 
  upper limit for the far-infrared luminosity of L$_{FIR}$=6$^.$10$^{10}L_\odot$,
  similar to the values found in starburst galaxies \citep{deutsch}.
  
  There is an absorption feature detected at 2.3 $\mu$m, but its shape does not 
  resemble a CO absorption band.
  Moreover the CO(3-1) and CO(4-2) absorption features are missing. 
  Probably high noise in this region hides the absorption.

  \subsection{HE 1029-1831}
  \label{apHE1029}
  
   \begin{figure}
   \centering
   \includegraphics[width=8.5cm]{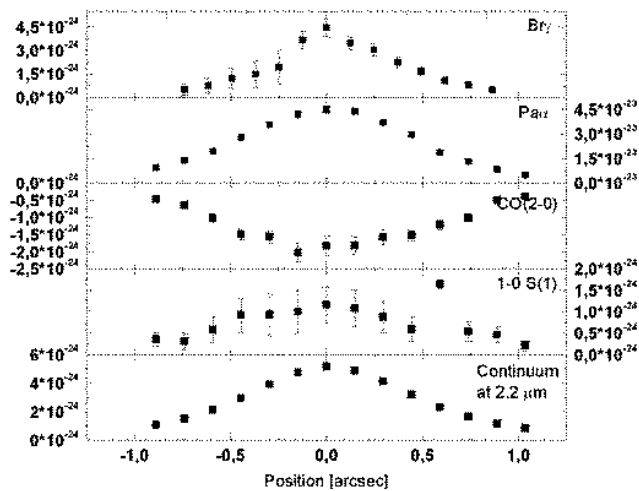}
      \caption{Flux of HE 1029-1831. 
      The flux of the Pa$\alpha$, Br$\gamma$, the (1-0)S(1) H$_2$-line
      and the flux deficit in ${\rm CO(2-0)}$ 
      is presented in units of W\,m$^{-2}$, the continuum flux is in units of 
      W\,m$^{-2}$\,$\mu$m$^{-1}$.}
      \label{HE1029flux}
   \end{figure}
  
  The spiral arms and the 
  bar structure are prominent features in the images, implying a face on
  view onto the galaxy.
  The flux ratios $[$NII$] \lambda$/H$\alpha$ and $[$OIII$]\lambda$/H$\beta$ indicate that HE 1029-1831 is an AGN. 
  Using flux ratios of $[$SII$]\lambda\lambda$6717.31/H$\alpha$ and $[$OIII$]\lambda$5007/H$\beta$, the galaxy is classified as HII/borderline galaxy, while the flux ratios of $[$OI$]\lambda$6300/H$\alpha$ $[$OIII$]\lambda$5007/H$\beta$ lead to a classification as AGN/borderline galaxy \citet{kewley}.
  In our spectrum, the Pa$\alpha$ line can be separated into a broad and a 
  narrow component with the broad component showing a width of 2081 km\,s$^{-1}$
  which is typical for a narrow line Seyfert 1 galaxy.
  This is consistent with the findings of \citet{nagao1} and \citet{rodriguez}
  who determined the FWHM of H$\alpha$ to be 1870 km\,s$^{-1}$ and classified the object as a narrow line Seyfert 1 galaxy, too.
  As in HE 1013-1947, the broad component's shape looks slightly asymmetric what could be caused by an unresolved helium-line.
  The Pa$\alpha$/Br$\gamma$ line ratio implies no significant reddening, which is consistent with the appearance of the continuum in Fig. \ref{spektren} as well as with the results of the photometry (see Fig. \ref{2color}).
  
  The equivalent width of the CO absorption feature decreases from (6 $\pm$ 3)~\AA~ at distances of 1\arcsec~ away from the nucleus to (4 $\pm$ 2)~\AA~ at the center.
  The small variations of the eqivalent width also substantiate only minor amounts of reddening.
  An equivalent width of 6 corresponds to equivalent widths found in early 
  K0-3 giants \citep{kleinmann}.
  
  HE 1029-1831 also shows extended emission in the H$_2$ 1-0S(1) transition with a total flux of (13.9 $\pm$ 1.4) $^.10^{-24}$\,W\,m$^{-2}$ (see Fig. \ref{HE1029flux}).
  The H$_2$ 1-0(S(1)/Br$\gamma$ line ratio is calculated as 0.29 $\pm$ 0.03.
  After \citet{rodriguez2}, a line ratio of 0.3 marks the transition from
  starburst galaxies to Seyferts, hence HE 1029-1831 shows only a comparatively
  small amount of activity for a Seyfert galaxy.
  In the 2-color diagram, the galaxy is located close to that of ordinary galaxies,
  revealing only small influences of the nucleus, too.
  
  HE 1029-1831 is also an IRAS source with F$_{12\mu m}$ = (139 $\pm$ 39) mJy, F$_{25\mu m}$ = (411 $\pm$ 40) mJy, F$_{60\mu m}$ = (2545 $\pm$ 15) mJy and F$_{100\mu m}$= (3704 $\pm$ 333) mJy. 
  This results in a far-infrared luminosity of L$_{FIR}$=1.9 $^.$10$^{11}$L$_\odot$.
  
  In summary, HE 1029-1831 probably is a luminous infrared galaxy with a weak narrow line AGN.

  \subsection{HE 1248-1356}
  \label{apHE1248}
  
  \begin{figure}
  \centering
  \includegraphics[width=8.5cm]{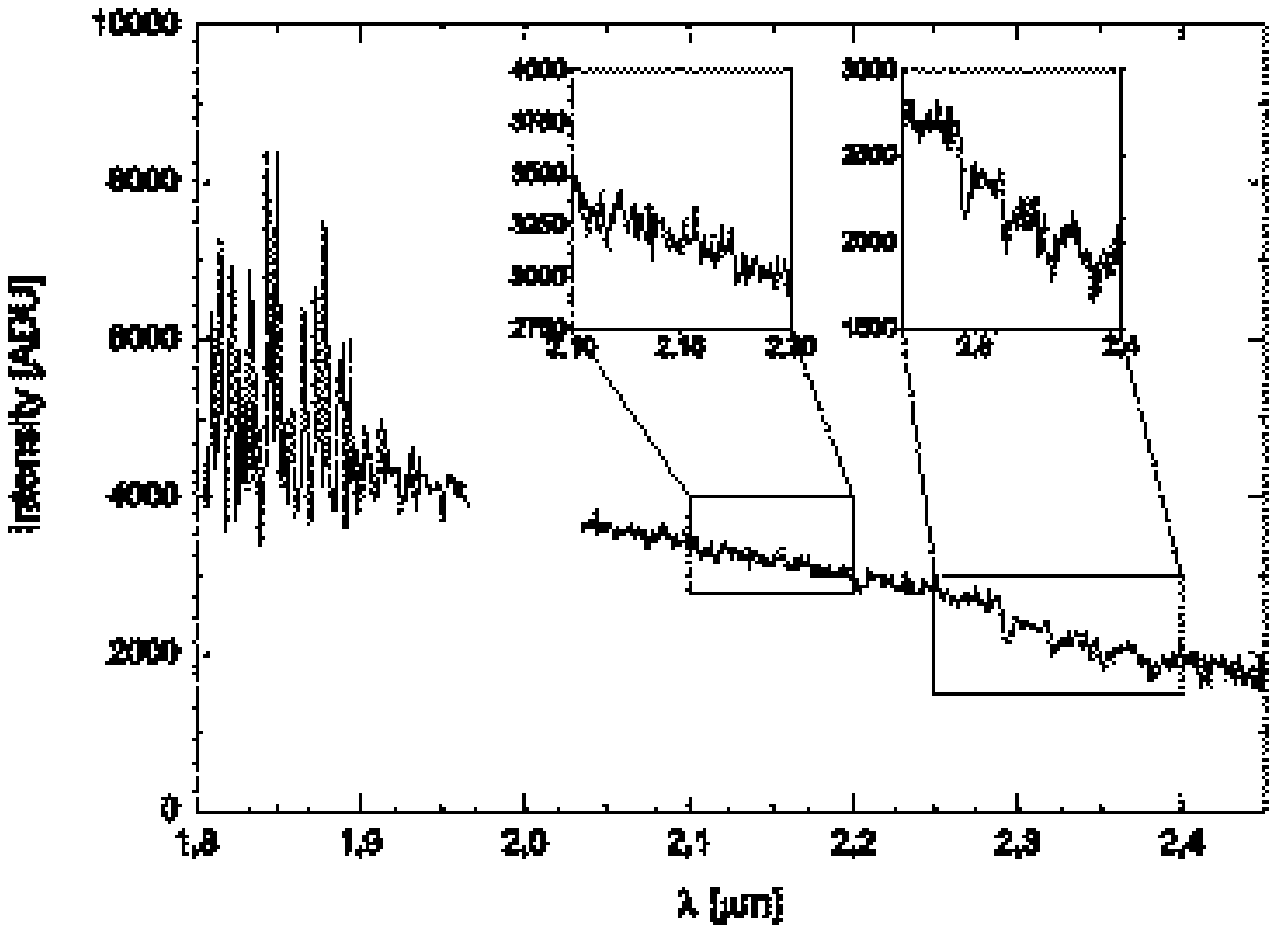}
      \caption{The over 5.2\arcsec (= 18.8 kpc) integrated spectrum of HE 1248-1356. 
      In addition to the overall K-band spectrum, the region of the Br$\gamma$ line as well as the CO absorption are shown.}
      \label{HE1248}
  \end{figure}
  
   \begin{figure}
   \centering
   \includegraphics[width=8.5cm]{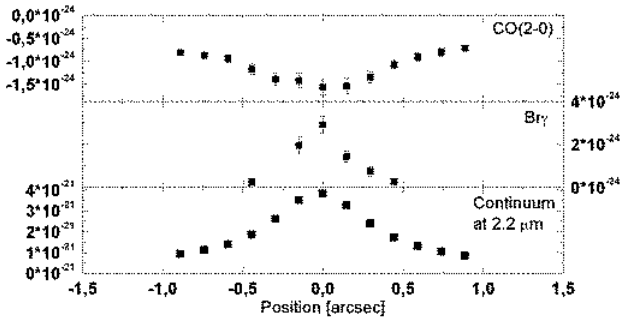}
      \caption{Flux of HE 1248-1356. 
      The flux of the Br$\gamma$
      line and the ${\rm CO(2-0)}$ flux deficit is given in units of W\,m$^{-2}$, 
      the continuum flux is in units of W\,m$^{-2}$\,$\mu$m$^{-1}$.}
      \label{HE1248flux}
   \end{figure}
  
  This galaxy is the closest one of the observed sources.
  It shows very prominent spiral arms at the west and east, the galaxy is inclined (i=63$\pm$6) and the velocity curve calculated from the ${\rm CO(2-0)}$ bandhead (Fig. \ref{rot+disp}b ) implies that the eastern arm is moving towards the observer.
  \citet{rodriguez} as well as \citet{maia} classified the object as a broad line Seyfert 1 galaxy.
  We do only detect a very faint but broad Br$\gamma$ line (Fig. \ref{HE1248flux}), though a weak Pa$\alpha$ line could be hidden in the noise produced by atmospheric absorption (see Fig. \ref{HE1248}).
  The CO absorption bands are very prominent features.
  The value of the CO-equivalent width at a distance of 1\arcsec~ to the center 
  is close to values typical for ongoing starformation.
  The depth decreases strongly towards the center, what indicates rising non-stellar
  continuum emission (Fig. \ref{HE1248flux}). 
  However, the slope of the continuum remains rather unaffected by this trend,
  the nuclear spectrum is already rather blue compared to the other observed
  sources.

  \subsection{HE 1328-2508}
  \label{apHE1328}
  
  The galaxy shows strong signs of interaction, an indication of a tidal tail can be seen, extending towards NE.
  The bright nucleus is not located at the center of the galaxy, the source at a distance of 3\arcsec SE of the nucleus is either a foreground star or a second nucleus.
  We tried to subtract a point like source from this possible second nucleus with the same process as for the other nuclei.
  The contribution could not completely be removed here, which supports the theory of a real second nucleus (or at least an extended object along the line of sight). 
  At 7\arcsec in the same direction, an additional source is located.
  This could be attributed to either a possible companion galaxy or a foreground star \citep[cf.][]{jahnke2}.
  No Hydrogen recombination lines are observed and the spectrum shows stellar CO absorption.
  In HE 1328-2508, however, the equivalent width of the CO(2-0) bandhead does not change with distance to the center as dramatically as it is observed in the other sources.
  In addition, the continuum slope shows no strong dependency on distance to the center.
  This could indicate only a small non-stellar contribution and is consistent
  with the location in the 2-color diagram near to the location of ordinary galaxies.
  The spectrum also shows molecular Hydrogen in emission at 2.122$\mu$m.
  Absorption makes the X-ray emission rather hard, its ROSAT hardness ratio 
  (HR1) is determined to 0.63 $\pm$ 0.03 \citep{fischerju}.

  \subsection{HE 1338-1423}
  \label{apHE1338}

  The ISAAC images resolve a clear bar structure and the host galaxy hence
  is classified as SB0 galaxy.
  \citet{bade} derived a spectral power-law index from the RASS of 1.41$\pm$0.29.
  In comparison to the continuum, HE 1338-1423 shows very strong recombination
  lines.
  The equivalent widths are (218$\pm$4.4)\AA~for Pa$\alpha$ and (18.1$\pm$0.1)\AA~
  for Br$\gamma$.
  The non-detection of CO-absorption, the strong change in continuum slope towards
  the center and the location in the 2-color diagram close to quasars all indicate to a strong non-stellar component.
  Our detected host galaxy colors are well in agreement with the results of \citet{jahnke}, who derived optical magnitudes of $B$=15.1, $V$=14.5, $R$=14.0 and $I$=13.4.
  
  \subsection{HE 2211-3903}
  \label{apHE2211}

  The galaxy shows a bar extending from NE to SW and very faint indications
  of two spiral arms.
  
  Despite a high S/N ratio of $\sim$45, the only feature resolved in the spectra is the Pa$\alpha$ line which shows a highly non-gaussian shape, implying complicated kinematics in this galaxy.
  Continuum slope changes are similar as in HE 1338-1423.
  \citet{maia} classified HE 2211-3903 as a Seyfert 1 galaxy via a broad H$\alpha$ line.
  
  The upper limits for the IRAS fluxes in the IRAS Vol. II catalog 
  (F60 is 0.77 Jy, F100 is an upper limit of 1.08 Jy) yield an 
  upper limit for the far infrared luminosity of L$_{FIR}$=5.2$^.$10$^{10}L_\odot$,
  similar to HE 1017-0305 and to the values found in starburst galaxies \citep{deutsch}.

  \subsection{VCV(2001) J204409.7-104324}
  \label{apVCVJ20}
  
   The host is an elliptical galaxy with a very bright nucleus.
   J204409.7-104324 is better known as Mrk 509. It is an IRAS source with fluxes
   F12=0.34 Jy, F25=0.74 Jy, F60=1.42 Jy and F100=1.43 Jy resulting in a
   L$_{FIR}$=0.9$^.$10$^{11}L_\odot$.
   
   Our colors are consistent with the results of \citet{glass2}, who classified Mrk 509 as Sy 1.5 galaxy and found average, de-reddened NIR colors of $J-H$=0.91, $H-K$=0.88 and $K-L$=1.35, significantly redder than colors of an ordinary in-active galaxy and consistent to our findings of a very strong non-stellar contribution. In their long term monitoring of the object, strong variations of $\Delta J$=0.65$^{mag}$ and $\Delta K$=0.62$^{mag}$ were detected.
   \citet{collins} find strong absorption in O$IV$ and column densities of multiple ionization stages of silicon (Si$II$, $III$, and $IV$) and Carbon ($II$, $III$, and $IV$) which are interpreted as a multiphase medium containing both collisionally ionized and photoionized gas in the bulge.
   
   Broad Pa$\alpha$ and Br$\gamma$ lines can be seen in the spectra.
   The broad component of the Pa$\alpha$ line shows a width of 2353 km\,s$^{-1}\pm$20\%, the width of the Br$\gamma$ line is estimated to a similar value of 2009 km\,s$^{-1} \pm$30\%, both being typical for a narrow line Seyfert 1 galaxy.
   The continuum slope indicates rather strong nuclear influences even at a distance of 2.4kpc from the nucleus, but the seeing during the observations was not very good with 2.6\arcsec.

\section{Summary and conclusions}  
   The Cologne Nearby QSO sample is well suited for spatially resolved studies of AGN. 
   The close distance of the objects allows for a photometric detection of the host galaxy even with only very short observation times.
   Low resolution spectroscopy of the stellar CO absorption band can give an upper limit to the central enclosed mass and first limited information on the dominating stellar population.
   The results of our observations can be summarized in the following points:
  
  \begin{enumerate}
  	\item{
	The dominating morphological class in this sample are disc dominated galaxies.
	Four of the nine hosts show a bar structure and spiral arms.
	Only one galaxy is found to be an elliptical galaxy.
	An underrepresentation of ellipticals is consistent with the results
	of other samples \citep[e.g.][]{smith,mcleodrieke,taylor,schade,jahnke}, 
	since most of the observed sources are lower luminosity AGN and the 
	probability to find an underlying disc-dominated host galaxy increases 
	with lower luminosity nuclei.
	}
	\item{
	In at least two galaxies, the appearance suggests that these 
	objects are interacting galaxies, in one galaxy a possible second nucleus
	is found.
	This supports the theory that nuclear activity may be triggered by
	merger events.
	}
  	\item{
	Seven of the nine galaxies show hydrogen recombination lines in
	either Pa$\alpha$, Br$\gamma$ or both.
	In three cases, the Pa$\alpha$ line shows a composition of a 
	broad and a narrow component, while in two cases only a broad
	component and in HE 0853-0126 only a narrow component is observed.
	For HE 2211-3903, the shape of the Pa$\alpha$ line points to
	more complicated kinematics.
	}
	\item{
	Three galaxies show extended molecular hydrogen emission in the 
	1-0S(1) transition.
	This appears to be a lower detection rate of Seyfert 1 galaxies with molecular hydrogen emission lines in comparison to the findings of other surveys \citep[e.g.][]{rodriguez2}.
	For significant conclusions, though, a larger sample size is needed.
	}
	\item{
	In five galaxies, stellar CO absorption is detected.
	With the exception of HE 1328-2508, all sources show a strong increase of the CO-equivalent width with growing distance to the center.
	In HE 1328-2508, HE 1029-1831 and HE 1013-1947, the CO(2-0)-equivalent width resembles the value found in ordinary elliptical or spiral galaxies. 
	In HE 1248-1356, the value can be associated to ongoing star formation.
	Only in the case of HE 1248-1356, the CO absorption allowed
	a determination of an upper limit to the central enclosed mass.
	}
	\item{
	The continuum slopes show a correlation to the detectability of the CO-absorption. In galaxies with significant reddening, the CO-absorption is diminished by the strong non-stellar continuum.
	}
	\item{
	In the $J-H$/$H-Ks$ 2-color diagram, the Seyfert galaxies are broadly 
	distributed
	over the region between normal galaxies and Quasars.
	}
	\item{
	The $H-Ks$ colors of the spiral hosts are typical for their non-active
	counterparts or slightly redder.
	}

  \end{enumerate}

\begin{acknowledgements}
We thank the anonymous referee for his valuable comments which helped improving the paper.

This work was supported in part by the Deutsche Forsch\-ungs\-ge\-sell\-schaft (DFG) via SFB494.
\end{acknowledgements}

\end{document}